\documentclass[11pt]{article}

\newcommand{\alps}	{\ensuremath{\alpha_S}}

\newcommand{\df}	{\ensuremath{\mathrm{d}}}
\newcommand{\e}		{\ensuremath{\mathrm{e}}}

\newcommand{\vc}	{\ensuremath{\mathbf{c}}}
\newcommand{\vb}	{\ensuremath{\mathbf{b}}}

\newcommand{\vr}	{\ensuremath{\mathbf{r}}}

\newcommand{\ap}	{\ensuremath{\alpha_{\mathcal{P}}}}
\newcommand{\apm}	{\ensuremath{(\ap-1)}}

\newcommand{\prg} {\texttt}

\begin{document}
\begin{titlepage}
\renewcommand{\thefootnote}{\fnsymbol{footnote}}
\begin{flushright}
Cavendish-HEP-95/07 \\ 
hep-ph/9601220 \\
January 1996
\end{flushright}
\vspace*{20mm}
 
\begin{center}
\textbf{\Large OEDIPUS: Onium Evolution, Dipole Interaction and
Perturbative Unitarisation Simulation\footnote{Research
supported by the UK Particle Physics and Astronomy Research Council}}
\vspace{10mm}
 
\textbf{G.P.~Salam} \\
\vspace{3mm}

\textit{Cavendish Laboratory, Cambridge University,} \\
\textit{Madingley Road, Cambridge CB3 0HE, UK} \\
\vspace{2mm}
e-mail: salam@hep.phy.cam.ac.uk
\end{center}
 
\vspace{20mm} 
\begin{abstract} 
OEDIPUS is a Monte Carlo simulation program which can be used to
determine the small-$x$ evolution of a heavy onium using Mueller's colour
dipole formulation, giving the full distribution of dipoles in rapidity
and impact parameter. Routines are also provided which calculate
onium-onium scattering amplitudes between individual pairs of onium
configurations, making it possible to establish the contribution of
multiple pomeron exchange terms to onium-onium scattering (the
unitarisation corrections). 
\end{abstract}

\end{titlepage}

\setcounter{footnote}{0}

\section*{PROGRAM SUMMARY}

\noindent \textit{Title of program:} OEDIPUS
\vspace{3mm}

\noindent \textit{Catalogue number:} 
\vspace{3mm}

\noindent \textit{Program obtainable from:} 
\prg{ftp://axpf.hep.phy.cam.ac.uk/pub/theory/oedipus.tar.gz} , see also
\prg{http://www.hep.phy.cam.ac.uk/theory/software/oedipus.html}
\vspace{3mm}

\noindent \textit{Licensing provisions:} None
\vspace{3mm}

\noindent \textit{Computers:} Tested on Dec Alpha, Sun
\vspace{3mm}

\noindent \textit{Operating system:} Unix (OSF3.2, SunOS-4.1.3,
Solaris-2.4) 
\vspace{3mm}

\noindent \textit{Program language used:} Fortran-90
\vspace{3mm}

\noindent \textit{Memory required to execute with typical data:} 1--10
MB 
\vspace{3mm}

\noindent \textit{No. of bits in a word:} 32/64
\vspace{3mm}

\noindent \textit{No. of lines in distributed program, including test
data, etc.:} 6186
\vspace{3mm}

\noindent \textit{Keywords:} Small-$x$; BFKL; Pomeron; Onium-onium
scattering; Dipoles; Monte Carlo; Unitarity.
\vspace{3mm}

\noindent \textit{Nature of the physical problem}

\noindent BFKL evolution [1] of a heavy (quark-)onium, to give full
information on the rapidity and transverse positions of gluons
carrying a small fraction $x$ of the onium's longitudinal
momentum. This information can be used to calculate a variety of
onium-onium scattering cross sections, including multiple pomeron
contributions, which restore unitarity. 
\vspace{3mm}

\noindent \textit{Method of solution}

\noindent A Monte Carlo simulation of the evolution in rapidity of the
gluon structure of the onium, using Mueller's colour dipole
formulation of small-$x$ physics [2].
\vspace{3mm}

\noindent \textit{Restrictions on the complexity of the problem}

\noindent The number of dipoles (gluons) in an onium grows as an
exponential of the rapidity, and inversely with the value of a cutoff
used to regulate an ultra-violet divergence. The time for evolution of
an onium is proportional to the number of dipoles, while that for
calculation of the onium-onium scattering amplitude goes as the square
of the number of dipoles. Memory restrictions also arise from the need
to store large configurations of dipoles (very large fluctuations can
occur in the number of dipoles present).
\vspace{3mm}

\noindent \textit{Typical running time} 

\noindent On a DEC Alpha 3000 processor, evolution generates about
10000 gluons per second. At a rapidity $Y=16$ and an ultra-violet
cutoff of $0.01$ times the onium size, a day's running will generate
about $10^5$ onium-onium scatterings.
\vspace{3mm}

\noindent \textit{References}

\noindent [1] Y.~Y. Balitski\v{\i} and L.~N. Lipatov, Sov. Phys. JETP
28 (1978) 822; \\
E.~A. Kuraev, L.~N. Lipatov, and V.~S. Fadin, Sov. Phys. JETP 45
(1977) 199; \\
L.~N. Lipatov, Sov. Phys. JETP 63 (1986) 904.

\noindent [2] A.~H. Mueller and B.~Patel, Nucl. Phys. B 425 (1994)
471; \\
A.~H. Mueller, Nucl. Phys. B 437 (1995) 107.

\vspace{1cm}

\noindent
{\large \textbf{LONG WRITE-UP}}

\section{Introduction}
There has been much interest recently in BFKL
\cite{BaLi78,KuLF77,Lipa86} type processes and their possible
observation at HERA and the Tevatron. Calculation of the cross
sections for a number of these processes, such as diffractive
dissociation (see e.g.\
\cite{BaLW95a,BiPe95b}) or exclusive vector meson production (e.g.\
\cite{BFLW96}) requires detailed information about the dominant
exchanged ``object'', known as the BFKL pomeron; for example one needs
an understanding of the triple pomeron vertex and of the transverse
distribution of small-$x$ gluons. At present, the only BFKL type
process which is fully calculable in perturbative QCD is high energy
onium-onium scattering. Mueller's colour dipole formulation of this
process \cite{Muel94a,MuPa94,Muel94b,Muel95} (for a related approach,
see \cite{NiZa93a,NiZZ94a}), offers a well defined way of performing
the necessary calculations, in the large $N_C$ approximation, where
$N_C$ is the number of colours.

The colour dipole formulation of small-$x$ evolution and high energy
onium-onium scattering is particularly suited to Monte Carlo
simulation, since the evolution is probabilistic in nature, and
because each branching (of one dipole into two) is independent of all
other branchings.

This paper is divided into three parts. The first gives a brief
overview of the small-$x$ dipole evolution of an onium and of the
issues relevant to a Monte Carlo simulation of such an evolution. The
second section examines the use of the dipole structures of a pair of
evolved onia to calculate onium-onium scattering amplitudes, including
the multiple pomeron exchange contributions which restore
unitarity. The final section gives a detailed description of the
structure and use of the OEDIPUS package.

\section{Dipole structure of an onium}
\subsection{Background}
One starts with an onium where the transverse separation between the
quark and anti-quark is $\vb_{01}$. The probability of generating a
gluon at a point $\vb_2$ carrying a small fraction $e^{-y}$ of the
light-cone momentum of the onium is \cite{Muel94a}:

\begin{equation}
   \frac{\df P}{\df y \df^2 \vb_2} = 
	e^{-y / \lambda}
	 \frac{\alps N_C}{2\pi^2} \frac{b_{01}^2}{b_{02}^2 b_{12}^2}.
\label{eq:trns_rte}
\end{equation}

\noindent $N_C$ is the number of colours ($N_C = 3$ for QCD),
$\alps$ is the strong coupling constant; $\lambda$, the effective
lifetime in rapidity ($y$) of the dipole, is related to the virtual
corrections (or more intuitively, conservation of probability):

\begin{equation}
	\lambda^{-1} =
	\frac{\alps N_C}{2\pi^2}  \int \df^2\vb_2 
	\frac{b_{01}^2}{b_{02}^2 b_{12}^2}.
\label{eq:virtual}
\end{equation}

\noindent In the limit of large $N_C$, the colour structure means that
the two new dipoles which arise ($\vb_{02}$ and $\vb_{12}$) can
themselves independently emit gluons, while the original colour dipole
is effectively destroyed. This branching of dipoles repeats itself
until the rapidity of any new gluons which would be produced exceeds
the maximum rapidity available. Through this branching process, one
can establish the probability of any given configuration of dipoles.

\subsection{Notes on implementation}
\label{sc:dipimp}
One of the main aspects of eq.~(\ref{eq:trns_rte}) is that it has
non-integrable divergences at $b_{02} = 0$ and $b_{12} = 0$. This
leads to the integral for $\lambda^{-1}$ being infinite. A solution
to this problem is to introduce a cutoff on the dipole size,
eliminating any region of $\vb_2$ where $b_{02} < \rho$ or $b_{12} <
\rho$. This has to be used in both eq.~(\ref{eq:trns_rte}) and
eq.~(\ref{eq:virtual}). The lifetime of a dipole of size $b$ is
therefore a function of $b/\rho$.

In the limit of small $\rho$ the effects of the cutoff should
disappear. However the number of small dipoles is large: the number of
dipoles of size $c$ from an onium of size $b$, after evolution through
$y$ is approximately:

\begin{equation}
        n^{(1)}(c,b,y) \df \log c \simeq \frac {b}{c \sqrt{\pi k y}}
        \exp(\apm y - \log(c/b)^2/ky) \df \log c.
\label{eq:n1}
\end{equation}

\noindent This is valid for $|\log(c/b)| \ll ky$. The BFKL power is
$\apm = 4 \log 2 \alps N_C / \pi$, and $k = 14\alps N_C \zeta(3)/\pi$,
with $\zeta(3) \simeq 1.202$ being the Riemann zeta function. The
smallest dipoles will be of size $c\sim \rho$, so as $\rho$ is lowered
the number of dipoles rises. In addition, the number of dipoles
increases exponentially with rapidity. The time taken to generate an
onium is proportional to the number of dipoles. The time to calculate
onium-onium interactions is proportional to the product of the number
of dipoles in the two onia.

A final complication will arise because there are very large
fluctuations in the numbers of dipoles \cite{Sala95} --- the
probability of obtaining a configuration with $n$ dipoles goes as
$\exp[-\pi (\log n)^2/4\alps N_C y]$, so that one has to allow for
configurations which are very much larger than the mean. Since
configurations generally need to be stored (for example to investigate
interactions between pairs of configurations), this can lead to
considerable memory consumption, especially at large rapidities.

The facility of imposing an upper limit on dipole sizes during the
evolution is also included, to allow a crude investigation into the
uncertainties due to infra-red effects. When the implementation of the
upper cutoff is turned on, both dipoles produced from a branching are
required to be smaller than the upper cutoff. The lifetimes of dipoles
are adjusted accordingly.

The evolution is carried out with fixed $\alps$, which is justified in
the limit of very heavy onia. There is no known unique way of
including a running coupling constant, however it would be possible to
modify the program to implement a scheme such as that used by Nikolaev
and Zakharov \cite{NiZa93a}.

\section{Onium-onium scattering}
\subsection{Background}
One wishes to obtain the amplitude $F(\vb, \vb', \vr, Y)$, for elastic
scattering between two onia of sizes (and orientations) $\vb$ and
$\vb'$, whose centres are separated by a transverse distance $\vr$,
with a total rapidity between them of $Y$ ($\simeq\log s$, where
$\sqrt{s}$ is the centre of mass energy).

Let $\gamma$ be a particular dipole configuration for an
onium, which contains dipoles of position and size $(\vr_1,\vc_1)
\ldots (\vr_{n_\gamma}, \vc_{n_\gamma})$. The interaction between two
such onia, moving in opposite directions is \cite{MuPa94, Muel94b,
Muel95}:

\begin{equation}
f_{\gamma, \gamma'} = \sum_{i = 1}^{n_{\gamma}}
        \sum_{j = 1}^{n_{\gamma'}}
        f(\vr_{i} - \vr_{j}', \vc_{i}, \vc_{j}'),
\label{eq:sumamps}
\end{equation}

\noindent where $f(\vr, \vc_{i}, \vc_{j}')$, the interaction between
dipoles of size $\vc$ and $\vc'$, whose centres are separated by $\vr$
is \cite{Muel94b,Sala95b}:

\begin{equation}
f(\vr, \vc, \vc') = \frac{\alps^2}{2} \left[ \log \frac{
         |\vr + \vc/2 - \vc'/2| |\vr - \vc/2 + \vc'/2| }
        {|\vr + \vc/2 + \vc'/2| |\vr - \vc/2 - \vc'/2| }
        \right]^2.
\label{eq:dipdip}
\end{equation}

\noindent One has to average $f_{\gamma, \gamma'}$ over dipole
configurations $\gamma$, $\gamma'$ of the two onia. Therefore at the
level of one pomeron exchange, the elastic amplitude is:

\begin{equation}
F^{(1)}(\vb,\vb',\vr,Y) = - \sum_{\gamma, \gamma'}
        P_{\gamma}(\vr_0, \vb, y)
        P_{\gamma'}(\vr_0 + \vr, \vb',  Y-y)
        f_{\gamma, \gamma'},
\label{eq:mcunit}
\end{equation}

\noindent where $P_\gamma(\vr_0, \vb, y)$ is the probability of
obtaining a dipole configuration $\gamma$ after evolution through $y$
of an onium of size $\vb$ centred at $\vr_0$. Note that in this case
the result can be shown to be independent of $y$, the division of
rapidity between the two onia (or equivalently of the longitudinal
frame in which the calculation is performed).

One can also calculate contributions to the amplitude involving the
exchange of $k$ pomerons \cite{Muel94b}:

\begin{equation}
F^{(k)}(\vb,\vb',\vr,Y) =  \frac{1}{k!}
        \sum_{\gamma, \gamma'}
        P_{\gamma}(\vr_0, \vb, y)
        P_{\gamma'}(\vr_0 + \vr, \vb',  Y-y)
        (-f_{\gamma, \gamma'})^k.
\end{equation}

\noindent These multi-pomeron terms, though formally non-leading in
$1/N_C$, are enhanced at high rapidities by the large numbers of
dipoles in each onium. All numbers of pomeron exchange can be resummed
to give an amplitude $F$ which explicitly satisfies the unitarity
bound:

\begin{equation}
F^{(k)}(\vb,\vb',\vr,Y) = -  \frac{1}{k!}
        \sum_{\gamma, \gamma'}
        P_{\gamma}(\vr_0, \vb, y)
        P_{\gamma'}(\vr_0 + \vr, \vb',  Y-y)
        (1 - \e^{-f_{\gamma, \gamma'}}).
\end{equation}

\noindent These last two equations are approximations based on there
being a large number of dipoles interacting. Multiple pomeron
contributions should be calculated in the frame $y = Y/2$, where
corrections that would arise from wave-function saturation, which is
not being calculated, are expected to be smallest \cite{Muel94b} (see
also \cite{MuSa96}).

\subsection{Implementation considerations}
\label{sc:ooimp}
One of the main difficulties in calculating onium-onium scattering
amplitudes is the large numbers of interactions which have to be
worked out: in principle the product of the numbers of dipoles in each
onium, multiplied by the number of points at which one needs to know
the interaction.

The first way to ease the problem is to note that the dipole-dipole
interaction, eq.~(\ref{eq:dipdip}), is relatively local: at large
distances it dies off as $1/r^4$ where $r$ is the separation between
the centres of the dipoles. Therefore it should be safe to neglect the
interaction between dipoles if their separation is sufficiently large,
say greater than twice the sum of their sizes. This is found to
reproduce the total dipole-dipole interaction to better than one
percent\footnote{Note that this may not correctly reproduce the
profile of the interaction at very large onium-onium separations,
where the density of dipoles is also dying off also as $1/r^4$ (though
enhanced by double leading logarithmic factors). However this region
is not important for the total interaction.}.

One generally wants to know the amplitude at many different impact
parameters (i.e.\ onium-onium separations), for example to work out
the total cross section (with the normalisation used above, just twice
the amplitude, integrated over all impact parameters). One way to do
this is to work out the interaction on a grid. Since there are large
fluctuations in the spatial extent of the dipole configurations, the
grid size has to be variable --- it should completely cover the area
where dipole-dipole interactions are significant. By summing the
interactions at each point of the grid one obtains an approximation to
the total amplitude for that configuration pair (the finer the grid,
the better the approximation). 

This approach is necessary if one wants to find a good approximation
to the total amplitude for each configuration pair, however it is slow
and does not offer any easy way to store the results so that they can
be retrieved for later analysis (say if one only wanted to look at 1
and 2 pomeron exchange, but after the run, decided that 3 pomeron
exchange would also be interesting) --- if one uses a grid which is
$100\times 100$, then one requires 40 kilobytes per pair of
configurations, for a single value of rapidity. To gain adequate
accuracy one needs several tens of thousands of events; with, say, 5
values of rapidity this gives several gigabytes of data.

One reason why it is important to sample a large number of
configuration pairs, is that rare, large configurations contribute
significantly to the total cross section \cite{Sala95b}. Multiple
pomeron contributions are dominated by rare, dense configurations. For
a given pair of dipole configurations, however, the variation of the
amplitude (at those points where it is significant) with impact
parameter is not so large. So it can be advantageous to evaluate fewer
points in impact parameter (i.e.\ making a worse approximation to the
amplitudes of individual configuration pairs) in exchange for a larger
sample of configurations. The limit of this is to evaluate the
onium-onium interaction, not on a grid, but along a radial line. In
addition, because there are fewer points in impact parameter it
becomes tractable to store the results. In fact the individual
amplitudes are not stored event by event, rather a probability
distribution for the amplitude is stored at consecutive bins in radial
position. This information can then be retrieved and used to determine
whichever quantity one is interested in.

\section{Structure and use of OEDIPUS}
The program comes in many different files in various directories:
\texttt{basic\_src/} contains all the common subroutines which are
involved with data input and output, initialisation and evolution of
onia, as well as routines for determining onium-onium interaction
amplitudes. It does not contain any code for analysing the results
from the evolution or interaction: this must be provided as a
``driver'' program by the user. A number of examples are to be found
in the directory \prg{samples/}, including drivers to examine the
spatial distribution of dipoles in the onium wave function and to
determine the total and elastic scattering cross sections.

\subsection{Storage of gluons and dipoles and evolution}
Gluons and dipoles each have a specific data type:

\begin{verbatim}
type gluon 
   complex(kind(1d0))    :: psn     ! stores the transverse position
   double precision      :: rapidity 
end type gluon 
\end{verbatim}

\noindent Complex variables are used throughout the program to deal
with transverse positions. The dipole type is

\begin{verbatim}
type dipole 
   type(gluon), pointer  ::  lo_y_gluon, hi_y_gluon 
   double precision      ::  size 
   type(dipole), pointer ::  child_1, child_2 
   integer               ::  index
end type dipole 
\end{verbatim}

\noindent It is useful to separate the gluon content of the dipole
into high rapidity and low rapidity, mainly for ease of coding. In the
case of the initial onium, the two quarks are represented by
\prg{gluon} types. The \prg{child\_1} and \prg{child\_2} pointers are
needed to store the tree structure of the evolved onium. If they are
not \prg{associated} then the dipole has no descendents (for example,
because any descendents would have a rapidity larger than the
maximum of the evolution). Each dipole in an onium has a different
\prg{index} --- this can be useful in analysing the branching
structure.

The following sequence of routines is needed to produce an onium and
evolve it, for each loop (or ``event'') of the program. First, various
internal counters must be reset by calling 

\begin{verbatim}
subroutine reset_counters
\end{verbatim}

\noindent Then an onium is produced with

\begin{verbatim}
subroutine init_onium(onium, onm_size, phi)
\end{verbatim}

\noindent This returns a \emph{pointer} \prg{onium} to a dipole 
of size \prg{onm\_size}, centred at $(0,0)$, with orientation
\prg{phi}. To produce a randomly oriented dipole, call

\begin{verbatim}
subroutine init_onium_rnd_phi(onium, onm_size)
\end{verbatim}

\noindent The onium is evolved up to a rapidity \prg{maxy} by calling 

\begin{verbatim}
subroutine evolve_onium(onium, maxy, n_tree)
\end{verbatim}

\noindent where, on return, \prg{n\_tree} is the total number of
dipoles in the tree. To be able to analyse the results, one wants only
the dipoles associated with a particular rapidity \prg{y}, so one must
extract them with:

\begin{verbatim}
subroutine extract_onium_dpls(onium, y, extrctd_dpls, n_extrctd)
\end{verbatim}

\noindent The array \prg{extrctd\_dpls(:)} is of type \prg{dpl\_pntr}
(arrays of pointers are not permitted in Fortran 90 --- the solution to
this is to define a type which contains nothing but a pointer), and
on return contains pointers to the \prg{n\_extrctd} dipoles. The array
must be sufficiently large: one can ensure that it is, by dynamically
allocating it to be of size \prg{(n\_tree / 2 + 1)}. The array of
dipole pointers can then be processed by the user.

This sequence allows one to first evolve the onium up to the maximum
rapidity that one is interested in, and then examine the dipole
structure over a range of rapidities.

Additional onia (for use simultaneously) can be produced, evolved, and
the dipoles extracted, by repeating the same set of calls (omitting
the call to \prg{reset\_counters}). 

\subsection{Onium-onium interaction}
\label{sc:ooprog}
Two sets of routines are provided which calculate amplitudes for
onium-onium scattering. The first returns a reasonable estimate for
the elastic scattering amplitude between a pair of onium
configurations, for all relevant impact parameters:

\begin{verbatim}
subroutine onm_onm_grid(ext_dpls1, n1, ext_dpls2, n2, grid, gh_lo, &
    & gh_int, gv_lo, gv_int) 
\end{verbatim}

It takes a pair of dipole sets (\prg{n1} dipoles in \prg{ext\_dpls1}
and \prg{n2} dipoles in \prg{ext\_dpls2}) and determines the range of
relative separations over which they interact. The onium-onium
interaction is then determined for each point of a grid covering this
area (\prg{grid(:,:)} must be a two-dimensional square array; its size
determines the resolution of the sampling) --- the
program returns information relating the array to the physical grid
being used: the horizontal and vertical positions of the bottom left
hand edge of the grid (\prg{gh\_lo} and \prg{gv\_lo}), as well as the
horizontal and vertical spacing (or interval) of the grid
(\prg{gh\_int} and \prg{gv\_int}). For each dipole pair, it only works
out the contribution to those points on the grid where the
dipole-dipole separation is not too large (as discussed in section
\ref{sc:ooimp}). The amplitudes are always worked out at the centres of
the grid rectangles.

The second approach works out a binned probability distribution for the
amplitude in many sections along a radial line. The details of the
binning are held in the \prg{binning} module which provides a variable
\prg{bn} of type \prg{bin\_prms}: 

\begin{verbatim}
      type bin_prms
         ! binning vars for amplitude prob distributions
         integer            :: n_f_lo, n_f_hi, nf
         double precision   :: f_mid, f_lo_lgint , f_hi_lgint 
         ! binning vars for r
         integer            :: n_r_lo,  n_r_hi, nr
         double precision   :: r_mid, r_hi_lgint
      end type bin_prms
\end{verbatim}

\noindent The binning of the amplitude probability distribution is
divided into two regions: the lower region is for amplitudes below
\prg{f\_mid} and the bins are indexed 0 to (\prg{n\_f\_lo - 1}),
logarithmically spaced with an interval \prg{f\_lo\_lgint}. It is
important to have a bin explicitly for the absence of interaction
(otherwise when integrating the interaction over a large area the
smallest bin amplitude could contribute significantly) --- this is bin
$-1$. The lowest non-zero bin must be sufficiently small to accurately
reproduce the integral of the dipole-dipole interaction (bearing in
mind that because the interaction dies off as $1/r^4$, there are
regions where the amplitude can be quite small, but whose overall
contribution to the amplitude is non-negligible). It is to allow
adequate coverage of this wide range of small amplitudes that
logarithmic binning (in the small amplitude region) is used. For
amplitudes above \prg{f\_mid} the binning is performed logarithmically
with an interval \prg{f\_hi\_lgint}. The large amplitude region will
contribute more significantly to the total amplitude and dividing the
binning into two regions allows one to sample it more accurately in
the regions which are more important.  The reason to have logarithmic
binning for large amplitudes is to be able to store the very large
amplitudes which can occur at large rapidities. It is sometimes useful
to have the index of the highest amplitude bin, which is \prg{nf =
n\_f\_lo + n\_f\_hi}. The variable \prg{bn} is initialised with
default values which reproduce total amplitudes to an accuracy of
about one percent.

The binning in $r$, the distance along the radius, is also divided
into two regions, but for different reasons: there are occasional
events with very large transverse extents. To be able to include them
with linearly spaced bins would require a prohibitively large number
of bins. So for large $r$, logarithmic spacing of the bins is
useful. But for small $r$, the most efficient sampling is one where
each bin corresponds to a region of similar area --- linear binning in
$r$ is more appropriate there. There are bins numbered 0 to
(\prg{n\_r\_lo - 1}) linearly spaced from $r = 0$ to
\prg{r\_mid}. Above this, there are \prg{n\_r\_hi}  logarithmically
spaced bins, with an interval \prg{r\_hi\_lgint}.

To help bin and ``unbin'' the data, the following functions are
provided: 

\begin{verbatim}
function bin_of_f(f)
\end{verbatim}

\noindent returns the bin number associated with an amplitude
\prg{f}. If the amplitude falls into a bin higher than \prg{nf}
then it is put into bin \prg{nf} --- this serves as a form of overflow
procedure: if the highest bin contains any entries then the program
should have been run with a larger number of amplitude bins. The
inverse function 

\begin{verbatim}
function f_of_bin(bin)
\end{verbatim}

\noindent returns the value of the amplitude corresponding to the
(logarithmic) centre of the bin (on average, the optimal value to use
in reconstructing amplitudes).

For a given dipole-dipole pair, it is useful to know to which radial
bins their interaction will contribute. The function 

\begin{verbatim}
function irad(r)
\end{verbatim}

\noindent aids this by returning the bin corresponding to a given
radius (the interaction region for a pair of dipoles is defined by two
radii, which are converted into the range of bins which must be
sampled). For each radial bin, it is useful to sample at a random
radius within the bin. The random value of $r$ returned by

\begin{verbatim}
function rad_rnd(i)
\end{verbatim}

\noindent has a probability distribution which increases linearly with
$r$ within the bin \prg{i}, replicating the distribution of radii that
would be obtained by choosing a random position in the corresponding
ring. Finally it can be necessary to know the lower radius of a given
bin (for example when analysing binned results), which is obtained by
calling

\begin{verbatim}
rad_lo(i)
\end{verbatim}

\noindent Given a pair of dipoles sets from evolved onia (\prg{n1}
dipoles in \prg{ext\_dpls1} and \prg{n2} dipoles in \prg{ext\_dpls2}),
the following routine

\begin{verbatim}
subroutine add_f_to_bins(ext_dpls1, n1, ext_dpls2, n2, fbins_single)
\end{verbatim}

\noindent adds $1$ to the appropriate amplitude bin of

\begin{verbatim}
fbins_single(-1:bn%nf, 0:bn%nr)
\end{verbatim}

\noindent for a radius chosen randomly with \prg{rad\_rnd} in each
radial bin.

\subsection{Data input and output, and initialisation}
A wide variety of information can be extracted from the evolved onium
and the program needs to have the flexibility to input and output
whatever information the user wants. In addition, it must be possible
to restart a Monte Carlo run to add to previously determined data ---
this should be as automatic as possible. A number of routines are
provided to make this simpler.

\begin{verbatim}
read_params_fdat(onm_size, maxy, n_prev_events, n_new_events)
\end{verbatim}

\noindent This routine first parses the command line. Defining
\prg{argn} to be the $n^{th}$ command line argument, it opens
\prg{arg1.dat} (a file containing formatted data from the evolution)
and \prg{arg1.prm} (a file containing the parameters of the
evolution). The number of evolutions (or events) to be performed in
this run of the program (\prg{n\_new\_events}) is the value of
\prg{arg2}. It then looks to see if \prg{arg3} is present: if not (or
if it is a dash) then the evolution is a continuation of a previous
one and the contents of \prg{arg1.prm} are read in, in the following
format:

\begin{verbatim}
jseed(1)  jseed(2)      
r_cut_lo  r_cut_hi        
onm_size  maxy
data_io
n_prev_events
\end{verbatim}

\noindent with the following definitions:

\begin{description}
\item[\prg{jseed(1:2)}] the pair of integers used as a seed for the
random number generator \prg{HWRGEN}, which uses l'Ecuyer's method,
described in \cite{Jame90}, as implemented in the HERWIG \cite{MWAK92}
event generator;
\item[\prg{r\_cut\_lo}] the lower cutoff on dipole sizes;
\item[\prg{r\_cut\_hi}] the upper cutoff (if it is negative, then no
upper cutoff is implemented);
\item[\prg{onm\_size}] the onium size;
\item[\prg{maxy}] the maximum rapidity up to which onia are evolved;
\item[\prg{data\_io}] a variable provided by the user in the module
\prg{data\_prms} holding any additional information needed by the
user's program for treating the data (e.g.\ a range of dipole
sizes being studied). It can be of any type (including user defined
type); 
\item[\prg{n\_prev\_events}] the number of ``events'' (i.e.\ Monte
Carlo evolutions) processed so far.
\end{description}

\noindent The routine also performs the initialisation which is
necessary before evolution can be carried out: a call to
\prg{init\_lives} which sets up an array tabulating dipole lifetimes as
a function of their size, and a call to \prg{nullify\_child\_set} which
nullifies pointers to child dipoles (since Fortran 90 has no mechanism
for specifying the initial status of a pointer). It is the user's
responsibility to read in the data from unit \prg{idat} (since the
format and nature of the data will vary from program to program).

If \prg{arg3} is present then the file \prg{starters/arg3} (path from
top directory of the distribution) is read in. The format is the
same as the first three lines of \prg{arg1.prm}. If either value of
\prg{jseed(1:2)} is zero then the default seed is used;
\prg{n\_prev\_events} is set to zero. It is then the user's
responsibility to initialise any data variables and provide default
values for \prg{data\_io}. This mechanism makes it much easier to run
the same program with different values of cutoffs and rapidity (the
most commonly varied parameter) by having a single set of starter
files in the the \prg{starters/} directory --- effectively by a
command line argument, rather than by having to edit a new parameter
file each time. A number of starter files are provided with the
distribution and are described in the file \prg{starters/README}. 

\begin{verbatim}
write_params_fdat(onm_size, maxy, n_events, param_arg, dat_arg)
\end{verbatim}

\noindent This routine is called to output the updated parameters of
the evolution (usually only the seed and the number of evolutions so
far, \prg{n\_events}, will have changed). It \prg{rewind}s both the
parameter and data devices. Normally the parameter and data devices
(\prg{param\_arg} and \prg{dat\_arg}) will be the same as those used
for input (\prg{iprm} and \prg{idat}) but the arguments are available
(though optional) because there is occasionally a need to use
different output devices (e.g.\ if the user wants to retain the
original data). It is the user's responsibility to output the data.

The advantage of using formatted data is that it is easy to examine or
plot. But it is very inefficient for storing large amounts of
data. Two routines are provided for dealing with unformatted data (the
parameters file is still formatted):

\begin{verbatim}
read_params_ufdt(onm_size, maxy, n_prev_events, n_new_events)
write_params_ufdt(onm_size, maxy, n_events, param_dev, ufdt_dev)
\end{verbatim}

\noindent They are very similar to the previous routines. The main
difference is that the data device is now \prg{iufdt}, which is
opened for unformatted output. Since the most common use of
unformatted input and output is for the binned probability
distribution of the amplitude (see section
\ref{sc:ooprog}) the parameter file contains the additional set of
parameters: \prg{bn} (a variable in the \prg{binning} module, of type
\prg{bin\_prms}) which defines the spacing of the
bins. With the unformatted data files it will often be necessary to
read them in again to analyse them. Because the various parameters may
be needed to know how to read in and process the data a routine

\begin{verbatim}
read_params_old_ufdt(prnt_size, maxy, n_events)
\end{verbatim}

\noindent is provided. It looks at \prg{arg1} to determine the files
to use and opens them (with the \prg{OLD} specifier, to prevent them
being accidentally altered). It does not perform any initialisation
for the evolution. 

For binned amplitude data, the format is fairly constant, so routines
are also provided for reading and writing the data:

\begin{verbatim}
subroutine allc_read_fbins(n_y, n_prev_events)
\end{verbatim}

\noindent This allocates an array 

\begin{verbatim}
fbins(-1:bn%nf, 0:bn%nr, n_y)
\end{verbatim}

\noindent in accord with the format for binned data discussed in
section \ref{sc:ooprog} reads it in from device \prg{iufdt}.  The
extra dimension comes in because one might want to store results for
several (\prg{n\_y}) values of rapidity from a single run. The reason
for the argument \prg{n\_prev\_events} is to produce the correct
normalisation for the data: on disk, it is stored so that each bin
contains the probability associated with it. When adding extra data,
it is convenient to use a normalisation where each bin contains the
number of events contributing to it. If \prg{n\_prev\_events} is 0
then the bins are not read in, but instead initialised to zero. If the 
argument is not present then the normalisation of the bins is not
changed.

To write the data, use the routine

\begin{verbatim}
subroutine write_fbins(n_events, iufdt_arg)
\end{verbatim}

\noindent It is sometimes useful to be able to specify a different
device, \prg{iufdt\_arg} (optional), for output than for input, for
example when writing out only a portion of the data. The number of
events needs to be provided so that the stored normalisation on output
is consistent with each bin containing a probability.

The data is stored as \prg{real} rather than \prg{double precision} to 
reduce the use of disk space.

\subsection{Some internal details}
The gluons and dipoles which are generated by evolution have to be
stored, for example so that one can then work out onium-onium
interactions. One of the main problems is that one doesn't know
beforehand how many dipoles will be produced --- the mean number of
dipoles can be estimated, however the fluctuations above this are very
large (see section \ref{sc:dipimp}). A solution is to dynamically
allocate memory for gluons and dipoles as the evolution proceeds. This
slows down the evolution (by nearly a factor of two on some systems)
because of the time needed for allocation and deallocation.

The alternative is to provide fixed size arrays for the dipoles and
gluons, where the size is chosen beforehand --- essentially by trial
and error. For cases where the largest configurations will be using a
significant fraction of a machine's resources, this has the
disadvantage that by having previously allocated the memory, it is
taken up for the whole duration of the job (even if there is nothing
stored there), whereas with dynamic allocation, the memory is used up
only when it is needed.

Limits have been coded into the \prg{gluon} module for the maximum
number of gluons: the variable \prg{max\_gluons}. For the situation
with dynamic allocation, this is the maximum number of gluons in each
onium. It is present simply to allow the program to handle cases which
might otherwise cause it to crash because they used all the available
data store for that process. In the case of the fixed arrays, the
quantity represents the maximum number of gluons in total (i.e.\ from
all evolved onia). Apart from that, when using the subroutines
described above for initialising onia, evolving them, etc., the
functionality should not depend at all on the method of gluon storage.


There are ways round the limitations on the maximum number of
dipoles. One could store very large configurations on disk (the
computer would normally be doing this anyway with the swap space ---
but if the program is responsible then the swap space is still left
for other processes). Alternatively, there are situations where the
dipole structure need not be stored at all. The occasions when a
stored dipole structure is needed is when looking at some kind of
correlation between dipole pairs (such as onium-onium interaction),
because for each dipole, one needs access to all the other
dipoles. But in these situations, the time per configuration is
normally the limiting factor, since this is proportional to the square
of the number of dipoles. In cases where the time taken for some
analysis is linear in the number of dipoles, often, it is not
necessary to have knowledge of the whole configuration. So an
evolution routine could be written which called an analysis routine
each time it reached the ``tip'' of a ``branch'' of a dipole ``tree'',
and it would never have to store more than a single path from the base
of the tree to the furthest ``tip''. 

\subsection{Compilation}
The program is provided in the form of a number of separate files, the
contents of which are documented in \prg{README} files. To aid in the
compilation process, two shell scripts are provided which select the
correct set of source files. Detailed information on their functioning
is provided as part of the distribution, however a summary is given
here:

\begin{verbatim}
mkf driver [ dyn | sml | lrg ]
\end{verbatim}

\noindent This is for compiling a driver routine which uses formatted
data i/o. The name of the driver should be provided without the
\prg{.f90} extension. The second command line argument indicates the
kind of gluon and dipole storage used --- the default is to use
dynamic allocation (corresponding to the option \prg{dyn}). Two fixed
size array storage options are also provided: \prg{sml} (up to 5000
gluons) and \prg{lrg} (up to 400000 gluons, which with a lower cutoff
of $0.01b$, should be adequate for evolution up to rapidities of
$y\simeq 15$). The executable is named \prg{driver\_dyn} (or \prg{sml}
or \prg{lrg} as appropriate) to allow multiple copies of the
executable, with different gluon storage, to coexist.

\begin{verbatim}
mku driver [ dyn | sml | lrg | noev ]
\end{verbatim}

\noindent This is for compilation of drivers which use unformatted
data i/o. It also compiles in the routines and modules associated with
binning of the amplitude. The command line arguments are the same as
for \prg{mkf}, except for the extra option available as the second
argument: \prg{noev}. This is to be used for drivers which analyse
data from a previous evolution --- no evolution routines are compiled,
and the data and parameter files are opened with the \prg{'OLD'}
specifier, meaning that they cannot be modified. Naming of the
executable follows the same convention as for \prg{mkf} except when
the \prg{noev} option is used, in which case the executable is named
\prg{driver} (since there should be no need to have multiple copies
compiled with different gluon storage methods).

\subsection{Test run}
The run described here tests the evolution of the onia, determination
and binning of the interaction amplitude, and subsequent analysis of
data for onium-onium interactions. It is for moderate rapidities and
very small statistics to ensure that it runs quickly (a few seconds)
on most systems. It is necessary to first compile the evolution
routine (to be found in the \prg{sample/} directory of the
distribution) with

\begin{verbatim}
> mku onm-onm-bnd dyn
\end{verbatim}

\noindent which will include in dynamic allocation of memory for
gluons and dipoles. The first command starts an evolution (with 10
events) where the maximum rapidity for each onium is 4, giving a total
rapidity for the collisions of 8. The lower cutoff used is $0.1$ times
the onium size. The binned, unformatted data are stored in
\prg{test\_y8.ufdt} (beware: this file is over 1MB in size) and the
parameters in \prg{test\_y8.prm}. The format of the variable
\prg{data\_io} (together with the defaults used here) is

\begin{verbatim}
data_io%y_wvfn = 0d0
data_io%y_int  = 0.5d0
data_io%n_y    = 5
\end{verbatim}

\noindent which specifies that the onium-onium interaction is
determined for 5 values of rapidity, each separated by $0.5$ (note
that this is for each onium --- so the intervals in total rapidity are
$1.0$). The maximum rapidity at which dipoles are extracted from each
onium is \prg{maxy}, but each onium is evolved to
\prg{data\_io\%y\_wvfn} (or \prg{maxy}, if this is larger) allowing
different runs to effectively use the same dipole configurations but
to examine the interactions at different values of rapidity. Various
messages are output at the start of the program, as different parts of
the startup procedure are accomplished. The line dealing with backups
may be preceded by other messages on some systems (e.g.\ those using
the NAG compiler --- see section \ref{sc:prtblty}), because the
compiler does not offer the facility of running a system command.

The next command causes the evolution to be continued for a further 5
``events'' (simply to allow a test of the continuation function). One
then wants to analyse the results of the evolution. In this test we
will look at the total amplitudes (the onium-onium amplitude
integrated over all impact parameters, which is equal to half the
onium-onium total cross-section). The program to do this is compiled
with

\begin{verbatim}
mku ftot noev
\end{verbatim}

\noindent The option \prg{noev} ensures that unnecessary routines
(such as those required for evolution) are not compiled in. The output
includes various messages at the beginning indicating the
initialisation procedure (these messages are sent to standard error,
while the numerical output is sent to standard output, to allow it to
be redirected to a file). The format of the numerical output from
\prg{noev} is one line for each value of rapidity, where each line
has:

\begin{verbatim}
(total rapidity)  (unitarised amplitude) (1 pomeron amplitude) ...
      (n_pom pomeron amplitude)
\end{verbatim}

\noindent (all on one line), and \prg{n\_pom = 4} is a parameter in
the program. It is the absolute values of the amplitudes which are
output.

A second test run, which tests the formatted data output is provided
and documented within the distribution. In addition, some other driver
and analysis routines are provided. These can all be found in the
\prg{samples/} directory.

Examples of results produced with OEDIPUS can be found in
\cite{Sala95,Sala95b}.

Typical rapidities for which adequate statistics are accessible with a
day's running are $Y \le 20$, using a lower cutoff of $0.01b$.

\subsection{Distribution}
The distribution (provided as a compressed tarred file) consists of a
number of directories. Details of their contents are documented in
extensive \prg{README} files. A shell script is provided to help
automate the installation procedure. Instructions are also included
for installation by hand.

\subsection{Portability}
\label{sc:prtblty}
The main issues of portability relate to accessing command line
arguments and causing a shell script (which performs backups of data
and parameter files) to be executed. Interfaces to the system
routines to perform these functions are provided as routines whose
names start with the \prg{lcl\_} prefix, held in a file

\begin{verbatim}
basic_src/common/lcl_???.f90
\end{verbatim}

\noindent where \prg{???} may be \prg{dec}, \prg{nag} or \prg{std} (or
the user can generate a new file for his/her operating system). The
file to be used must be specified as part of the installation
procedure. With the NAG compiler (version 2.1), there does not seem to
be any way of causing a shell script to be executed (the usual unix
\prg{system} subroutine seems to be absent). The routines for
accessing command line arguments while available, do not seem to be
documented. If the Fortran 90 compiler offers no mechanism for
accessing the command line arguments then one can use
\prg{lcl\_std.f90}, which prompts the user for the information held in
each of the arguments. It conforms to the standard.

Local implementations are also necessary for writing large,
unformatted arrays to disk (on some systems, the i/o buffer is not
sufficiently large and the array must be written in small
sections). Details are provided in the \prg{lcl\_???.f90} files.

To port to a non unix system (such as VMS), it would in addition be
necessary to rewrite the compilation scripts.

\subsection{Global data}
All global data is stored in the form of modules. These are summarised
in table~\ref{tbl:modules} (numbers in brackets refer to default
values). 
\begin{table}[p]
\begin{tabular}{|l|l|p{3.5in}|} \hline
Module name & Variables & Description \\ \hline
\prg{constants} & \prg{alpha\_s} & $\alps$, the strong coupling
constant ($0.17777\ldots$)\\ \cline{2-3}
 & \prg{nc} & $N_C$, the number of colours in QCD ($3$)\\ \cline{2-3}
 & \prg{alpha\_no\_nc} & the value of $\alps$ to be used when it
\newcommand{\ttt} {\textttt}appears without a factor of $N_C$
(=\prg{alpha\_s}). 
\\ \hline
cuts & \prg{r\_cut\_lo} & the lower cutoff \\ \cline{2-3}
 & \prg{r\_cut\_hi} & the upper cutoff \\ \cline{2-3}
 & \prg{hi\_cut\_implem} & true if an upper cutoff is being used \\
\hline
\prg{iomodule} & \prg{iprm} & i/o device for parameters (10)\\
\cline{2-3} 
 & \prg{idat} & i/o device for formatted output (12)\\ \cline{2-3}
 & \prg{iufdt} & i/o device for unformatted data (14)\\ \hline
\prg{gluons} & \prg{type gluon} & the type definition for gluons \\ 
\cline{2-3}
 & \prg{type dipole} & the type definition for dipoles \\ \cline{2-3}
 & \prg{type dpl\_pntr} & type definition for a pointer to a dipole
\\ \cline{2-3}
 & \prg{max\_gluons} & the maximum number of gluons to be generated
\\ \hline
\prg{data\_defs} & \prg{data\_io} & variable of user chosen type, with
details of data i/o. This module must be provided by the user. 
\\ \hline
\prg{sub\_defs} & & interface blocks for all the main subroutines 
\\ \hline
\end{tabular}
\caption{A list of the modules (and their main contents) needed by
user's programs}
\label{tbl:modules}
\end{table}

\noindent In addition there are modules associated with the
binning routines. These are summarised in table~\ref{tbl:binning}.

\begin{table}[p]
\begin{tabular}{|l|l|p{3.5in}|} \hline
\prg{binning} & \prg{type bin\_prms} & type definition to store all
the parameters for binning of amplitudes \\ \cline{2-3}
& \prg{bn} & variable actually containing the information \\
\hline
\prg{fbins\_mdl} & \prg{fbins} & array containing amplitude bins
\\ \hline
\prg{bin\_interface} & & interface blocks for the binning routines 
\\ \hline
\end{tabular}
\caption{Modules associated with routines for binning the amplitude
probability distribution}
\label{tbl:binning}
\end{table}

\section*{Acknowledgements}
I would like to thanks B.R.~Webber and A.H.~Mueller for suggesting
this work, as well as for many useful comments. In addition,
M.H. Seymour's introduction to Monte Carlo simulation techniques (in
\cite{Seym92}) has been very helpful during the development of this
program.

\clearpage


\newpage

\noindent \textbf{TEST RUN}

\noindent The input and output of the test run are shown below:

\begin{verbatim}
> onm-onm-bnd_dyn test_y8 10 y4
 Starting a new evolution
 HWRGEN will use its default seed
 Lifetimes have been initialised
 Have allocated bins
If there are no error messages then the backups have been done
 Am about to evolve
> onm-onm-bnd_dyn test_y8 5
 Continuing a previous evolution....
 HWRGEN has been initialised
 Lifetimes have been initialised
 Have allocated bins
If there are no error messages then the backups have been done
 Am about to evolve
> ftot test_y8 
 Reading from a previous evolution....
 HWRGEN has been initialised
 Have allocated bins
 4.0000E+00 1.3183E-01 1.3466E-01 2.9205E-03 9.6577E-05 3.3521E-06
 5.0000E+00 1.8955E-01 2.0176E-01 1.4294E-02 2.4481E-03 4.1440E-04
 6.0000E+00 2.0235E-01 2.1490E-01 1.4465E-02 2.2114E-03 3.4580E-04
 7.0000E+00 4.2093E-01 4.7515E-01 6.3966E-02 1.1441E-02 1.9750E-03
 8.0000E+00 4.4738E-01 5.0385E-01 6.5855E-02 1.0901E-02 1.7452E-03
>
\end{verbatim}

\noindent The starter routine \prg{y4} is the following (default seed, 
lower cutoff of $0.1$, no upper cutoff ($-2.0$), onium size of
$1.0$, and maximum rapidity for each onium of $4.0$):

\begin{verbatim}
 0      0
 0.1    -2.0
 1.0    4.0
\end{verbatim}

\end{document}